\documentclass[aps,preprint]{revtex4-1}

\usepackage{graphicx}
\usepackage{epstopdf}
\usepackage{bm}
\usepackage{lineno}

\usepackage[utf8]{inputenc}
\usepackage{amsmath,amssymb,lmodern} 

\newcommand{\be}{\begin{equation}}
\newcommand{\ee}{\end{equation}}
\newcommand{\nn}{\mbox{} \nonumber \\ \mbox{} }
\newcommand{\ba}{\begin{eqnarray}}
\newcommand{\ea}{\end{eqnarray}}

\newcommand{\E}{{\bf E}}
\newcommand{\B}{{\bf B}}
\newcommand{\J}{{\bf J}}
\newcommand{\A}{{\bf A}}

\renewcommand{\div}{{\rm \,div\,}}

\newcommand{\Bf}{{magnetic field}}

\newcommand{\NS}{neutron star}
\newcommand{\ms}{magnetosphere}
\newcommand{\NSs}{{neutron stars}}

\newcommand{\Ef}{{electric  field}}

\newcommand{\EM}{electromagnetic}
\newcommand{\BH}{{black hole}}
\newcommand{\BHs}{{black holes}}
\newcommand{\Sc}{Schwarzschild}
\newcommand{\mss}{magnetospheres}

\newcommand\lo{\mathrel{\raise.3ex\hbox{$<$}\mkern-14mu\lower0.6ex\hbox{$\sim$}}}
\newcommand\go{\mathrel{\raise.3ex\hbox{$>$}\mkern-14mu\lower0.6ex\hbox{$\sim$}}}

\begin{document}
\title{Electromagnetic draping of  merging neutron stars}

\author{Maxim Lyutikov\\
Department of Physics and Astronomy, Purdue University, \\
 525 Northwestern Avenue,
West Lafayette, IN
47907-2036 }

\begin{abstract}
We first derive a  set of equations describing general stationary configurations of relativistic  force-free plasma, without assuming any geometric symmetries. We then demonstrate that 
 electromagnetic  interaction of  merging neutron stars is necessarily dissipative due to the effect of electromagnetic draping -  creation of dissipative regions near the star (in the single-magnetized  case) or at the magnetospheric boundary (in the double-magnetized case). Our results indicate that even in the single magnetized case we expect  that relativistic jets (or ``tongues'') are   produced, with correspondingly beamed emission pattern.
\end{abstract}

\maketitle

\section{Introduction}

The detection of gravitational waves associated with a short  GRB \citep{2017PhRvL.119p1101A} identifies merger of \NSs\ as  the central engine. It is highly desirable to detect any possible precursor to the main event.
\cite{2001MNRAS.322..695H} \citep[see also][]{2012ApJ...757L...3L} argued that magnetospheric interaction during double \NS\ (DNS) merger can lead to the production of \EM\ radiation. The  underlying mechanism  advocated in those works is a creation of inductive \Ef\ due to the relative  motion of \NSs. Both singly magnetized (1M-DNS) and double magnetized case (2M-DNS) are possible \citep{2019MNRAS.483.2766L}.
The 1M-DNS case is similar to the Io-Jupiter interaction \citep{1969ApJ...156...59G}.  Other relevant works include \cite{2011Natur.478...82N,2020ApJ...893L...6M,2021ApJ...923...13C}. 

Similarly to the DNS merger, in the case of merging \BHs, or BH-NS mergers, motion of the \BH\ through \Bf\  (generated  ether by the accretion disk or through \NS\ \ms) 
leads to generation of inductively-induced outflows, even by  a non-rotating  \Sc\ \BH\ \citep{2011PhRvD..83l4035L,2011PhRvD..83f4001L,2010Sci...329..927P,2012ApJ...754...36A}.

The approach taken by \cite{2001MNRAS.322..695H,2019MNRAS.483.2766L}, heuristically, follows that of \cite{GJ}, in that a quasi-vacuum approximation is used at first. This leads to the generation of dissipative regions,  pair production and ensuing nearly-ideal plasma dynamics. 
Resulting charges and currents modify the magnetospheric structure. 
 In the axisymmetric case this leads to the pulsar equation \citep{1973ApJ...182..951S,BeskinBook}.   The pulsar equation, a variant of the Grad-Shafranov equation \citep{Grad1967,Shafranov1966}, is a scalar equation for axially-symmetric relativistic   force-free configurations.  Axial symmetry allows  introduction of  an associated Euler potential, which, together with the $\div \B=0$ and ideal conditions reduce the force-balance to a  single scalar equation.

 In the case of merger double neutron stars systems, there is no geometrical  symmetry that can be used to reduce the force-balance to a single equation.
In this paper we first derive  equation  governing relativistic   force-free configurations {\it without assuming axially symmetry}, \S \ref{Relativisticforcefree}.  It is a set of two nonlinear elliptic  equations for two  Euler potential, with initially unknown dependence of the electric potential.
It turns out to be prohibitively complicated. 

In \S \ref{expansion2} we take an alternative approach: expansion in small \Ef\ (small  velocity). We demonstrate that the \EM\ fields ``pile-up'' near the surface of the \NS, creating regions with large \Ef. Similar effects occur in 2M-DNS scenario, \S \ref{Doublem}.

\section{Relativistic force-free configurations}
\label{Relativisticforcefree}

First we derive a  set of equations describing general stationary configurations of relativistic  force-free plasma, without assuming any geometric symmetries.

Let us represent the  \Bf\ in terms of Euler  potentials $\alpha-\beta$, and stationary  \Ef\ in terms of the electrostatic potential $\Phi$ (factors of $4\pi$ are absorbed into definitions of fields)
\ba &&
\B = \nabla \alpha \times  \nabla \beta
\nn &&
\E = -\nabla \Phi
\ea
Ideal condition 
\be
\E \cdot \B = \nabla \Phi \cdot( \nabla \alpha \times  \nabla \beta)
\ee
requires  $ \Phi(\alpha )$ or  $ \Phi(\beta )$. For definiteness let's assume  $ \Phi(\alpha )$.  This is an initially unknown function that needs to be found as part of the solution with given boundary conditions.

Also, we  impose  orthogonality  condition 
\be
 (\nabla \alpha \cdot \nabla \beta) =0
 \label{orth} 
 \ee
   Then vectors $\nabla \alpha, \, \nabla \beta, $ and $ \nabla \Phi$ form an orthogonal triad. Surfaces of constant $\alpha, \, \beta, \,   \Phi$ are mutually orthogonal.

 Force balance
\be
\Delta \Phi \nabla \Phi +( \nabla\times \B )\times \B=0
\label{balance}
\ee
takes the form
\ba &&
\nabla \beta \left( \nabla \alpha \cdot(- \nabla \alpha \Delta \beta+  {\cal L} (\alpha,\beta))  \right) +
\nn &&
\nabla \alpha \left(   (\nabla \alpha \cdot \nabla \alpha) \Phi^\prime \Phi^{\prime \prime} + \Delta \alpha  (\Phi^\prime )^2 - 
\left( \nabla \beta \cdot ({\cal L} (\alpha,\beta)  + \Delta \alpha  \nabla \beta )  \right) 
 \right) =0
 \label{EQQ}
 \ea
where 
\be
{{\cal {L}} } (\alpha,\beta) \equiv (\nabla \alpha \cdot \nabla) \nabla \beta -(\nabla \beta \cdot \nabla) \nabla \alpha
\ee
and primes denote $  \Phi^\prime = \partial_\alpha  \Phi(\alpha)$. 

Both terms  in (\ref{EQQ}) should be zero independently 
\ba &&
\nabla \alpha \cdot(- \nabla \alpha \Delta \beta+  {\cal L} (\alpha,\beta)) 
\label{01}
\\ &&
 (\nabla \alpha)^2 \Phi^\prime \Phi^{\prime \prime} + \Delta \alpha  (\Phi^\prime )^2 =
\left( \nabla \beta \cdot ({\cal L} (\alpha,\beta)  + \Delta \alpha  \nabla \beta )  \right) 
\label{2}
 \ea
Equations (\ref{01}) - (\ref{2}), together with constraint  (\ref{orth}) represent two equations for two Euler potentials $\alpha$ and $\beta$. 

Some further modifications can be done. 
Eq. (\ref{01}) can be written as 
\be
\nabla \alpha \Delta \beta =  {\cal L} (\alpha,\beta) +g \nabla \beta
\label{1}
\ee
 where $g$ is an arbitrary function. Scalar product  (\ref{1}) with $\nabla \beta$ gives
\be
(\nabla \beta \cdot {\cal L} (\alpha,\beta)) = -  g (\nabla \beta \cdot \nabla \beta)
\ee
 Eq.   (\ref{2}) then becomes
 \be
 (\nabla \alpha)^2    \Phi^\prime \Phi^{\prime \prime} + \Delta \alpha  (\Phi^\prime )^2 =  (\nabla \beta)^2 (\Delta \alpha -g)=0
  \ee
  Or
  \be
   \Delta \alpha  \left(  (\Phi^\prime )^2 -   (\nabla \beta)^2 \right) +  (\nabla \alpha)^2  \Phi^\prime \Phi^{\prime \prime} + g   (\nabla \beta)^2=0
   \label{eq2}
   \ee
Equations (\ref{1})  and (\ref{eq2})  can be used instead of (\ref{01}) - (\ref{2}). Function $g$ should be chosen to fit the boundary conditions. A simple example is considered in Appendix  \ref{Static}.

The set of equations  (\ref{01}) - (\ref{2}) - (\ref{orth}) or (\ref{1}) -  (\ref{eq2}) - (\ref{orth})  describe general relativistic force-free equilibrium. It's a nonlinear set of equations for two functions  $\alpha$ and $\beta$  with initially unknown  $ \Phi(\alpha )$.

\section{Metal  sphere moving through force-free \Bf}
\label{expansion2}

\subsection{Boundary conditions}

\begin{figure} [h!]
    \includegraphics[width=.99\linewidth]{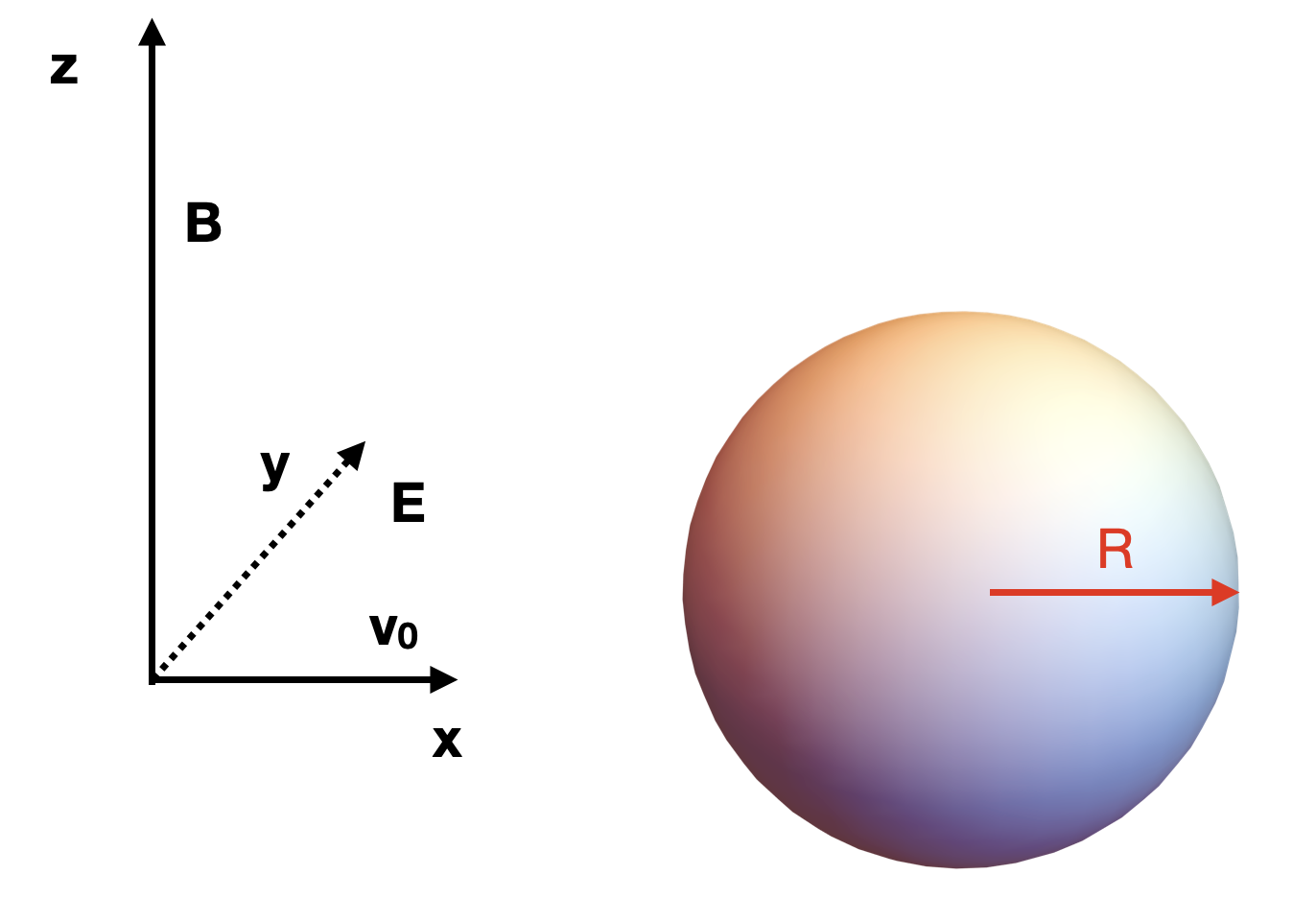}
\caption{Geometry of the system.  In the frame of the sphere at large $x \to - \infty$ the \Bf\ is along $z$ axis, \Ef\ is along $y$ axis, so that plasma is moving in positive $z$ direction with velocity  $v_0$. } 
\label{coords}
\end {figure}

Let in the frame of a conducting ball the \Bf\ at infinity be  along $z$ and \Ef\ is along  $y$ axis (so  that \EM\ velocity is along $x$), see Fig. \ref{coords}.  The \Bf\ is assumed to be non-penetrating, the ball is unmagnetized.

A set of equations that needs to be solved is
\ba && 
\rho_e \E + \J \times \B=0
\nn &&
\div  \J =0
\nn &&
\E \cdot \B =0
\ea 
force balance, stationarity and ideality. Boundary conditions are
\ba && 
B_z(x=-\infty) =B_0 
\nn && 
E_y(x=-\infty) =v_0 B_z
\nn && 
{\bf e}_r \cdot  \B  |_{r=R}  =0
\nn &&
{\bf e}_r \times \E |_{r=R} =0
\ea
The last two imply no normal \Bf\ and no tangential \Ef\  on the surface.

In terms of Euler potentials
\ba &&
\beta (x= - \infty) = -x
\nn &&
\alpha (x= - \infty) = B_0 y
\nn &&
{\bf e}_r \cdot  ( \nabla \alpha \times \nabla \beta )  |_{r=R}  =0
\nn &&
{\bf e}_r \times \nabla \alpha |_{r=R} =0
\ea
 (Landau gauge for \Bf\  is better at $x=\infty$). Thus, at  $x=-\infty$ we have  $ \Phi = - v_0 B_0 y = - v_0 \alpha$. 

The resulting system of nonlinear elliptical equations with unknown $\Phi(\alpha)$  turns out to be prohibitively complicated, hence we have to resort to approximate methods - expansion in terms of the velocity $v_0$.

  \subsection{Metal ball in static \Bf, $v_0 =0$}
  \label{Static}


As a zeroth-order, we start with conducting ball in external \Bf.
      In this case \Bf\ and vector potential are  a sum of constant vertical field $\B_v$ and dipole field $\B_d$
  \ba &&
     \B_0 = \B_v+\B_d=  \left\{ \left(1-{R^3}/{r^3} \right) \cos \theta ,
     - \left( 1+ \frac{R^3}{2 r^3}   \right)  \sin \theta , 0\right\} B_0
     \nn &&
     \A= \left\{0,0,1-{R^3}/{r^3} \right \}  r \sin\theta B_0/2
     \nn &&
     \B_v = \left\{ \cos \theta ,
     - \sin \theta , 0  \right\} B_0
     \nn &&
\B_d =      \left\{ -{R^3}/{r^3}  \cos \theta ,
     -  \frac{R^3}{2 r^3}   \sin \theta ,  0 \right\} B_0
     \label{confinedd}
\ea
     
    Euler potentials  are
     \ba &&
  \alpha_0 = \frac{1}{2} B_0 \sin ^2\theta  \left(r^2-\frac{R^3}{r}\right)
\nn &&
  \beta_0 =\phi 
\ea

Scalar magnetic potential
\be
          \Phi_B =  \left(1+ \frac{R^3}{2r^3}\right) r \cos \theta B_0
          \label{PhiB} 
\ee
so that
   $\B_0 = \nabla \alpha_0 \times \nabla \beta_0 = \nabla \Phi_B$.
 
 Importantly,
 \be
 (\nabla \alpha_0) \cdot (\nabla \Phi_B) =0
 \ee
 
\begin{figure} [h!]
    \includegraphics[width=.99\linewidth]{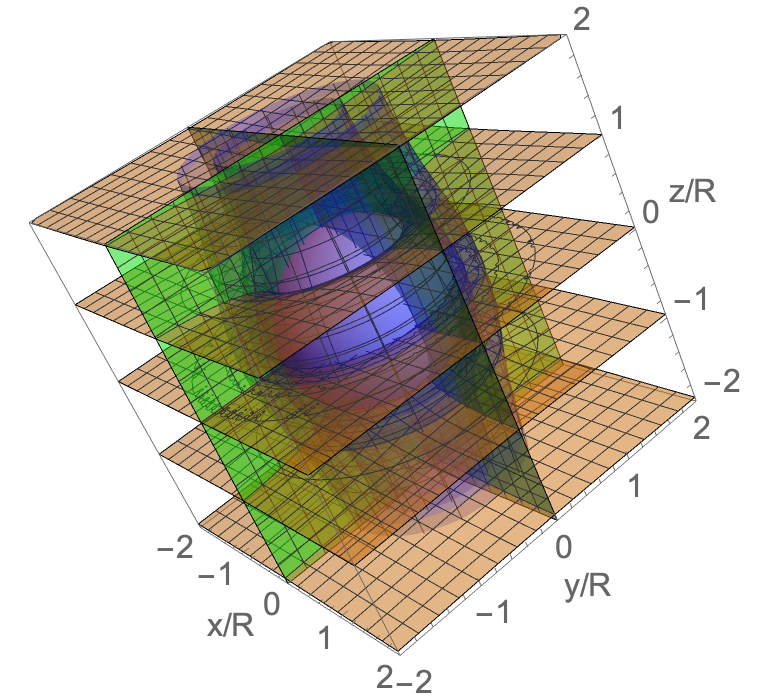}
\caption{Orthogonal surfaces of constant  $  \alpha_0,\,   \beta_0 $ and  $\Phi_B$.} 
\label{orthogonal}
\end {figure}

 Thus, Euler potentials
 $\alpha_0,  \beta_0 $ and $\Phi_0$ form a  mutually orthogonal triad of surfaces, see Fig. \ref{orthogonal}
 \be
 \nabla   \alpha_0 \perp  \nabla    \beta_0 \perp  \nabla  \Phi_B
 \ee
 
We find
\ba && 
  \nabla \alpha_0= \left\{   \left(r+ \frac{R^3}{2r^2} \right) \sin ^2\theta ,\frac{  \left(r^3-R^3\right) \sin  \theta  \cos  \theta }{r^2},0\right\} B_0
\nn &&
   \nabla \beta_0= \left\{0,0,\frac{1 }{r \sin  \theta }\right\}
  \nn &&
   (\nabla \alpha_0 \cdot \nabla \beta_0) =0
   \nn &&
    {\cal L} = \frac{B_0   \left(-4 r^3+3 R^3 \cos (2 \theta )+R^3\right)}{2 \sin  \theta   r^4}  {\bf e}_\phi
    \nn &&
      {\cal L} \cdot  \nabla \alpha_0=0
    \nn &&
      {\cal L} \cdot  \nabla \beta_0 = \frac{B_0  \left(-4 r^3+3 R^3 \cos (2 \theta )+R^3\right)}{2 \sin ^2 \theta  r^5}
      \nn &&
      \Delta \beta_0 =0
     \nn &&
     (\nabla \beta_0)^2= \frac{1}{ \sin ^2 \theta  r^2}
     \nn &&
      (\nabla \alpha_0)^2= \frac{B_0^2 \sin ^2\theta  \left(3 R^3 \cos (2 \theta ) \left(R^3-4 r^3\right)-4 r^3
   R^3+8 r^6+5 R^6\right)}{8 r^4}
   \nn &&
   \Delta \alpha_0 = -\frac{B_0 \left(-4 r^3+3 R^3 \cos (2 \theta )+R^3\right)}{2 r^3}
   \ea

Eq. (\ref{1}) becomes
\ba &&
{\cal L} (\alpha,\beta) +g_0  \nabla \beta_0=0
\nn &&
g_0= \frac{B_0 \left(4 r^3-3 R^3 \cos (2 \theta )-R^3\right)}{2 r^3} 
   \ea
 
 For $\Phi=0$ Eq. (\ref{eq2}) becomes
\be
g_0 = \Delta \alpha_0
\ee
And it is indeed satisfied.

  \subsection{First order expansion in $v_0$}
  \label{Firstorder} 
  
In Appendix \ref{vanishing} we demonstrate that in  first order expansion  in $v_0$ the surfaces of constant $\alpha-\beta-\Phi$ remain unchanged.

Let's expand the force balance  (\ref{balance}) for small velocity  $v_0 \ll 1$.
In the zeroth order 
  \be
  \nabla\times \B _0=0
 \ee

We expect that electric  potential is first order in $v_0$
\ba &&
 \Phi \propto  v_0 \sim \frac{E}{B_0}
 \nn &&
 \Phi \propto {\cal{O}} (\epsilon)
\ea

The key point is that  the  force balance is second order in $v_0$:
 \ba &&
\Delta \Phi \nabla \Phi +( \nabla\times \delta \B )\times \B_0=0
\nn && 
\Delta \Phi \nabla \Phi\propto {\cal{O}} (\epsilon^2)
\nn && 
( \nabla\times \delta \B )\propto {\cal{O}} (\epsilon^2),
\label{101}
\ea 
while the constraint
\be
\E \cdot \B \propto \E \cdot \B_0 \propto {\cal{O}} (\epsilon)
\ee
is first order.
Thus, if we are limited to terms linear in $v_0$, we need to consider only the constraint: the force balance is violated only in  $v_0^2$. 

For \Bf\ $\B_0$ let's use the magnetic potential (\ref{PhiB}). 
Then we need to find $\Phi$ such that 
\be
(\nabla \Phi) \cdot (\nabla \Phi_B) =0
\ee
Clearly  any 
\be
 \Phi(\alpha_0) f (\phi)
 \ee
 satisfies this condition. 
 
 At $x=-\infty$ the electric potential is
 \be
 \Phi = - y v_0 B_0 = - r \sin \theta \sin \phi v_0 B_0
 \ee
 Thus,
 \ba &&
 \Phi(\alpha_0) = -  v_0 B_0 \sqrt{2 \alpha_0}
 \nn &&
  f (\phi) = \sin  \phi
  \ea
  And finally
 \ba &&
 \Phi =  -\sqrt{1-{R^3}/{r^3}}   \times r  \sin \theta \sin \phi \,   v_0 B_0
 \nn && 
 \E = - \nabla  \Phi  = 
 \left\{\frac{\left(1+ \frac{R^3}{2 r^3}\right) }{\sqrt{1-{R^3}/{r^3}}}  \sin \theta  \sin \phi    , \sqrt{1-{R^3}/{r^3}}  \cos \theta  \sin (\phi
   ),\sqrt{1-{R^3}/{r^3}} \cos \phi \right\}v_0 B_0 \to
   \nn &&
   \left\{\frac{\sqrt{3} \sqrt{R} \sin  \theta  \sin \phi}{2 \sqrt{\delta
   _r}},\frac{\sqrt{3} \cos  \theta  \sqrt{\delta _r} \sin \phi
   }{\sqrt{R}},\frac{\sqrt{3} \sqrt{\delta _r} \cos \phi}{\sqrt{R}}\right\} v_0 B_0
   \nn &&
   \delta_r = r-R
   \label{EE1}
   \ea
    By construction $\E \cdot \B_0 =0$. The radial component of the  \Ef\ diverges - this is the \EM\ draping.   Also, in  Appendix \ref{comp}  we compare \Ef\ (\ref{EE1}) with other relevant cases.

\begin{figure} [h!]
    \includegraphics[width=.32\linewidth]{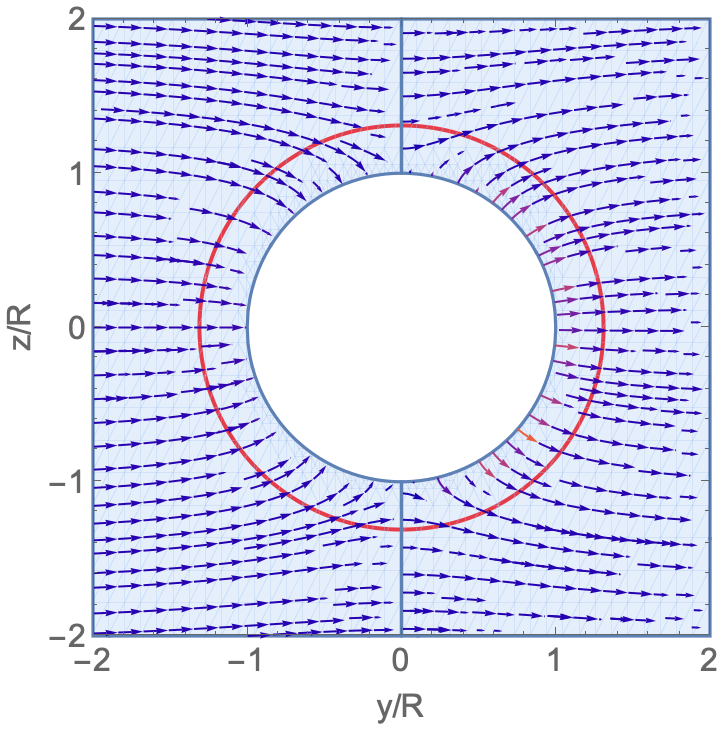}
\includegraphics[width=.32\linewidth]{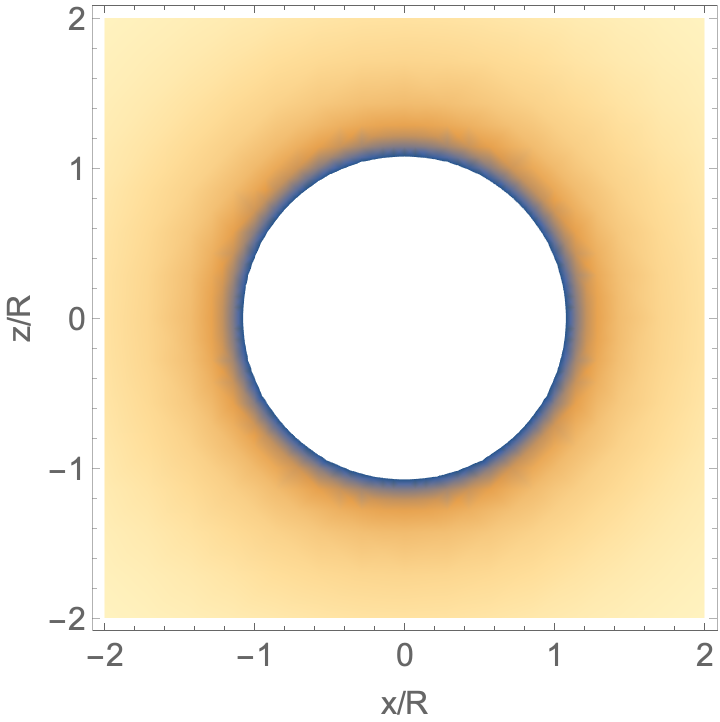}
\includegraphics[width=.32\linewidth]{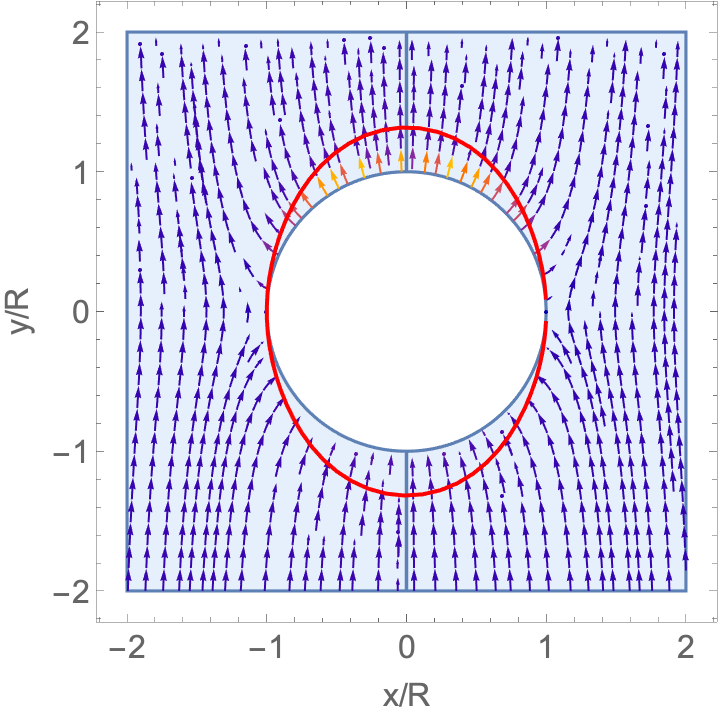}
\caption{Electric field (\ref{EE1}) in the $x=0, \, y=0$ and $ z=0$ planes (for $y=0$ the \Ef\  is  $  \sqrt{ (x^2+z^2)^{3/2} -1}/ (x^2+z^2)^{3/4}  {\bf e}_y v_0 B_0$. Red lines indicate regions where $\beta_{EM} =1$ ($v_0 =0.75 $ is assumed for plotting) } 
\label{outline1}
\end {figure}

\begin{figure}  [h!]
    \includegraphics[width=.9\linewidth]{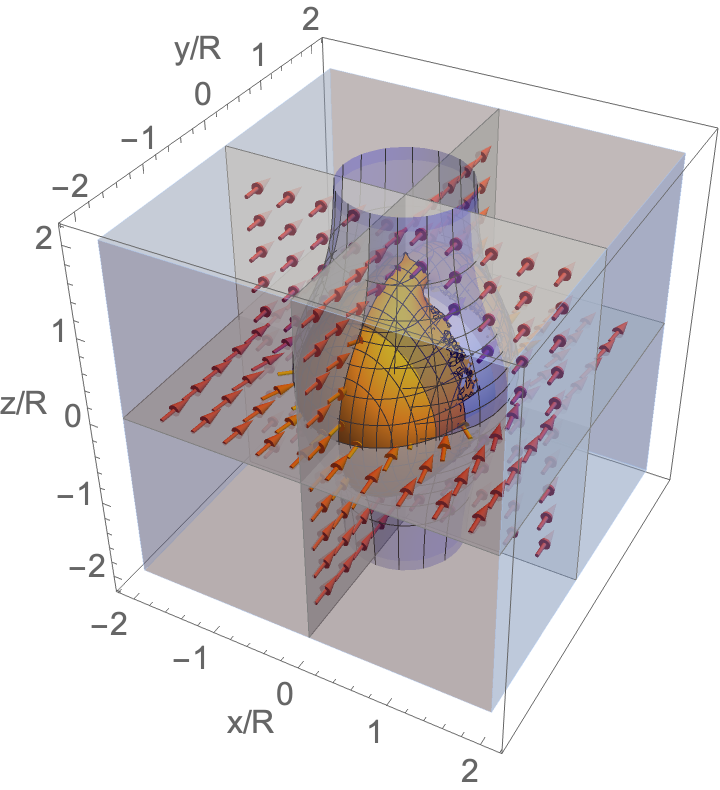}
\caption{3D view of first order \Ef\ (\ref{EE1}).  The central sphere is the \NS. Blue surface is the \Bf\ flux surface (\Bf\ lines lie on the surface pointing in the $z$ direction. Arrows are \Ef\ sliced at $x=0, y=0, z=0$. In the frame of the \NS\  plasma is moving in the $+x$ direction. Bounded ear-like surfaces are regions where $\beta_{EM} $ becomes larger than 1. } 
\label{outline2}
\end {figure}

\begin{figure}  [h!]
    \includegraphics[width=.49\linewidth]{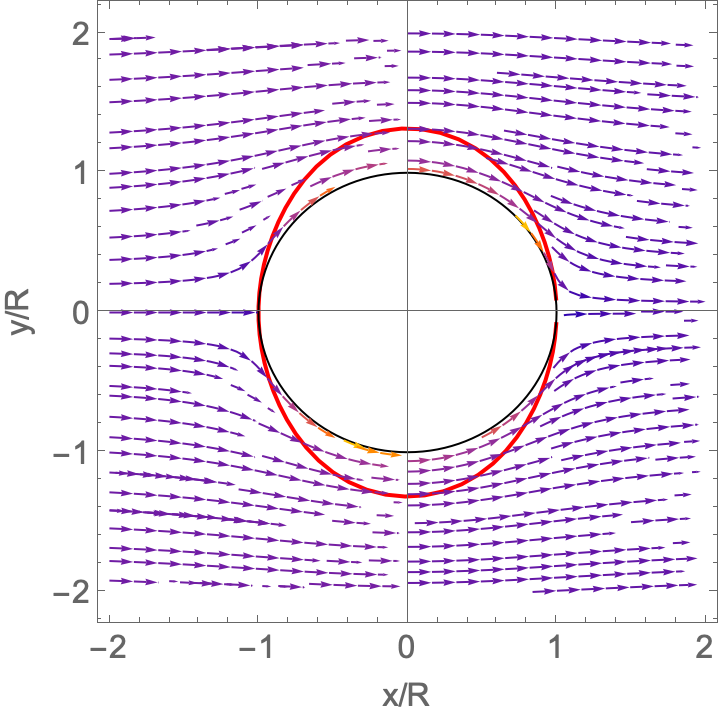}
\includegraphics[width=.49\linewidth]{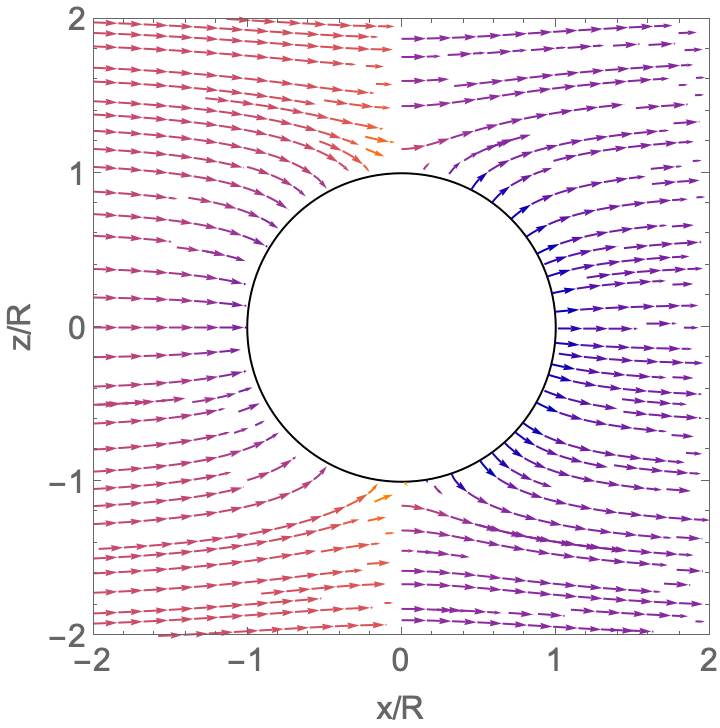}
\caption{Flow lines in the $z=0$ and $y=0$ plane.  A slight disconnection at $x=0$ is an artifact of the plotting procedure. In the plane $x=0$ the velocity is  $\beta_{EM} = (y^2+z^2)  ^{3/4}/ \sqrt{(y^2+z^2)  ^{3/2}-1} v_0 {\bf e}_x$. Red lines indicate regions where $\beta_{EM} \geq 1$. } 
\label{outline3}
\end {figure}

\begin{figure}  [h!]
    \includegraphics[width=.99\linewidth]{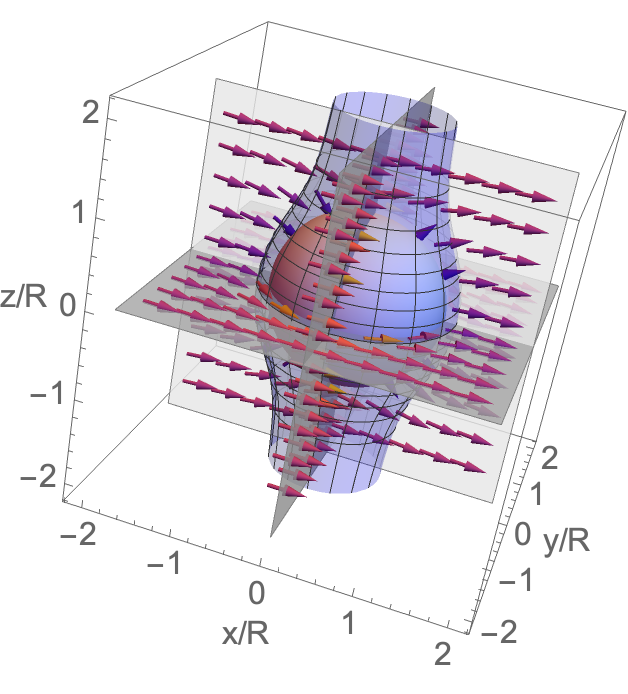}
\caption{Velocity plot.} 
\label{outline4}
\end {figure}

    Given the \Ef\ (\ref{EE1}),
          the induced charger density is
           \be
           \rho_e = \div \E = -\frac{9 R^6 \sin  \theta  \sin \phi}{4  r^7
   \left(1-{R^3}/{r^3}\right)^{3/2}} B_0 v_0 \to -\frac{\sqrt{3} \sqrt{R} \sin  \theta  \sin \phi}{4 \delta _r^{3/2}} B_0 v_0
   \label{rhoe}
   \ee
The      
   electromagnetic velocity is
   \ba &&
   {\bf v}_{EM} = \frac{\E \times \B_0}{\B_0^2} 
   \nn &&
v_r = \frac{2 r^3 \left(2 r^3+R^3\right) \sqrt{1-{R^3}/{r^3}}}{4 \left(r^3-R^3\right)^2-3
   R^3  \left(R^3-4 r^3\right) \sin ^2\theta }  \sin \theta \cos \phi 
\nn && 
v_\theta=
\frac{4 r^3 \left(r^3-R^3\right) \sqrt{1-{R^3}/{r^3}}}{4 \left(r^3-R^3\right)^2-3 R^3
    \left(R^3-4 r^3\right) \sin ^2\theta  } \cos \theta \cos \phi 
\nn &&
v_\phi =  -\frac{\sin \phi   }{\sqrt{1- R^3/r^3}}\to -\frac{v_0 \sin \phi}{\sqrt{3} \sqrt{\delta _r/R}} 
\label{vEM} 
     \ea
     see Figs. \ref{outline1}-\ref{outline4}.

     The condition
     $ {\bf v}_{EM} =1$ is satisfied at approximately
     \be
     \frac{\delta r}{R}  =\sin^2 \phi  \frac{v_0^2}{3} 
     \label{betaEM1}
     \ee
     This is the estimate of the thickness and location of the draping layer. It is maximal at the plane $x=0$ ($\phi = \pi/2$).    In the $\phi=\pi/2$ plane ($x=0$) the condition $\beta_{EM}= 1 $  is satisfied at 
     \ba && 
    \frac{  r_{EM} }{R} = \gamma_0^{1/3}
    \nn &&
    \gamma_0 = \frac{1}{\sqrt{1-v_0^2}}
    \ea

    \subsection{Second order in $v_0$}
    
    As we discussed above, the first order perturbations come not from the dynamics, but from the constraint $\E \cdot \B=0$. We can then use the $\propto v_0$ terms to construct the second order expansion.

     The charge density (\ref{rhoe}) and the  electromagnetic velocity (\ref{vEM}) 
     lead  to the appearance of charge-separated current 
     \be
     {\bf J}_{EM} = \rho_e {\bf v}_{EM} 
     \label{JEM1}
     \ee
     (it is of $ {\cal{O}} (\epsilon^2)$ order).   The current $ {\bf J}_{EM}$ is not the total current, only its transverse charge-separated  part, see below.
     Naturally, 
        \be
    \rho_e \E+  {\bf J}_{EM}  \times \B_0 =0
    \ee

     The  most radially-divergent  $\phi$-component can be easily found
     \be
    {\bf J}^{(2)} _\phi =\frac{9
   \sin  \theta  \sin ^2\phi}{4 r \left(r^3-R^3\right)^2}  \times R^6 v_0^2 B_0
   \label{Jphi} 
\ee
     Near $r\to R$
     \be
 {\bf J}_{EM}       \approx \left\{-\frac{\sin (2 \phi)}{4 \delta _r},0, \left(  \frac{R}{4
   \delta _r^2}-\frac{3 }{4 \delta _r} \right)  \times \sin  \theta  \sin ^2\phi \right\} B_0 v_0^2
   \label{JEM}
     \ee
     The toroidal current increases the most.  Since largest gradients are in radial direction, that leads to growth of $B_\theta$, see Eq. (\ref{Btheta}).


      Using (\ref{Jphi}), neglecting $B_r$ component (small near the surface, non-penetrating \Bf),
    we
    find divergent terms 
    \be
    B_\theta = \left(-\frac{3 R^2}{4
   \left(r^3-R^3\right)} +\frac{\ln \left(\frac{r^2+r R+R^2}{(r-R)^2}\right)}{4 r}  \right)  \sin  \theta  \sin ^2\phi R B_0 v_0^2
   \label{Btheta}
   \ee
       
    The most dominant  divergent term is
    \be
    \delta B_\theta=  - \frac{R}{4 \delta _r } B_0 v_0^2 \sin  \theta  \sin ^2\phi
    \label{deltaBtheta} 
    \ee 
Eq.  (\ref{deltaBtheta})  gives an estimate of the \Bf\ perturbation - hence the justification of the first order expansion. The condition 
    $ \delta B_\theta \leq B_0$ implies that the first order expansion is valid for
    \be
 \frac{   \delta_r}{R}  \geq v_0^2,
 \ee
    consistent  with (\ref{betaEM1}). 

Thus, both the \Ef\ and the \Bf\ diverge on the surface - this is \EM\ draping. The \Ef\ diverges in linear terms  in $v_0$, \Bf\  in  $v_0^2$.
The ratio of divergent terms in the first order \Ef\ and second order \Bf\ is
\be
\frac{E_r} {  \delta B_\theta} = 
-2  \sqrt{3}  \frac{ \sqrt{ \delta r /R} }{v_0  \sin \phi }
\label{E1overB2}
\ee
Thus, the  divergent second order term in \Bf\  cannot generally compensate for the divergent first order term in \Ef.

    Next, the longitudinal current 
    \be
    {\bf J}_{\parallel} = G(r,\theta, \Phi)  \B_0
    \ee
    follows from stationary condition
    \be
    \div (  {\bf J}_{EM}+  {\bf J}_{\parallel}) =0
    \ee
    We find
    \be
     \div{\bf J}_{EM} \approx  \left( \frac{1}{\delta _r^2} - 3 \frac{1}{R \delta _r} \right)  \sin \phi \cos  \phi v_0^2 B_0
     \label{divJEM}
     \ee
    Function $G$ must be $\propto \sin \phi \cos  \phi$, and we find
    \be
     \div ( G \B_0) \approx  \left( 2 \delta_r  \cos \theta \partial_r G - \sin \theta \partial_\theta G \right)  \frac{3}{4} \sin (2 \phi) \frac{B_0}{R}
     \ee
     
     To match $\theta$-independent  $\div{\bf J}_{EM}$ (\ref{divJEM}) function $G$ should be necessarily divergent either at $\theta =0 $ (the  $\partial_\theta G$ term)
     or at $\theta = \pi/2$ (the  $\partial_r G$ term)

  \section{Double magnetized (anti)aligned case} 
  \label{Doublem} 
  
\begin{figure} [h!]
    \includegraphics[width=.99\linewidth]{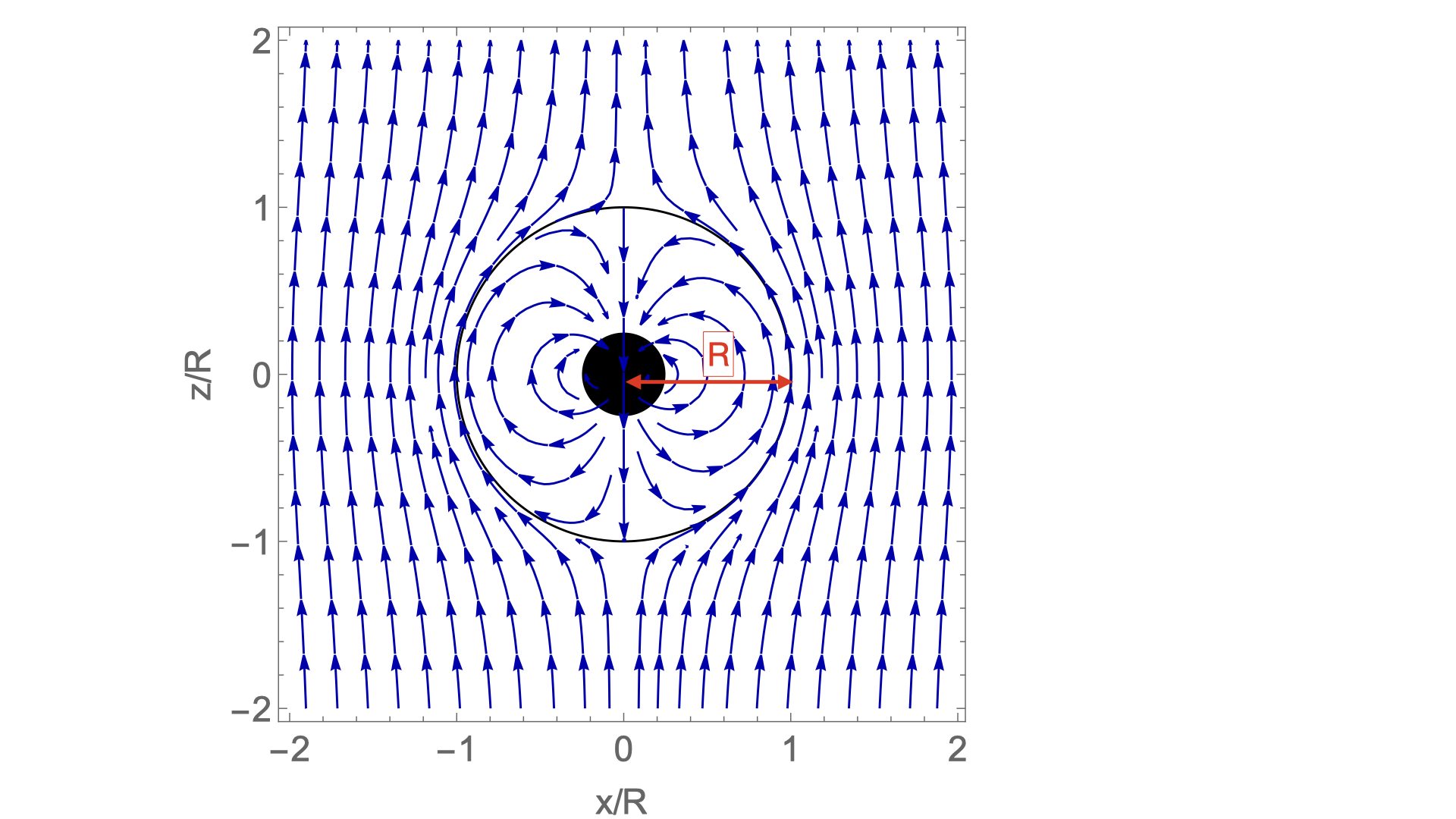}
\caption{Double magnetized anti-aligned case.  Topologically disconnected  intrinsic dipolar field matches the external field at $r=R$. The black circle in the center indicates the \NS.} 
\label{confined1}
\end {figure}

  Results of the single magnetized \NS\ can be generalized to the double magnetized aligned or anti-aligned  case in the case when the reconnection effects are not important and the \mss\ remain topologically disconnected \citep[see][for the case when the \mss\ are strongly coupled]{2020ApJ...893L...6M,2021ApJ...923...13C}.
     Recall that for a metal ball in  external \Bf, the field is  a sum of dipole and external field. For double magnetized case, then the parameter $R$ is the radius where the field of the star matches the external field, Fig. \ref{confined1}. Equivalently, in expression for $\B_d$, a change  $R^3 B_0 \to \mu$ in Eq. (\ref{confinedd}), the magnetic moment of the star.  In the anti-aligned case, when the magnetic moment opposes the external field, there are no currents; in the opposite aligned case there is a toroidal  surface current at  $R$. 
     
The location of the boundary between the external \Bf\ and that of the \NS\ \ms\ is not fixed now (for single-magnetized case it was the surface of the star). But as we discuss in Appendix \ref{vanishing} any distortion of the surfaces is second order in $v_0$. Thus, in the linear regime all the  previous derivations for the 1M-DNS case remains valid.

\section{Discussion}

In this paper we argue that effects of \EM\ draping - creation of dissipative layer near the merging \NS\ may  lead to  generation of observable  precursor emission. The draping effect is well known in space and astrophysical plasmas  \citep{2004AIPC..719..381C,2006MNRAS.373...73L,2008ApJ...677..993D}. In the conventional MHD limit, when the \Ef\ is not an independent variable, creation of the magnetized layer (for super-Alfvenic motion) does not lead to dissipation, only break-down of the weak-field approximation in the draping layer. 

We argue that relativistic plasmas are different. In this case the s \Ef\ is an independent dynamic variable; also charge densities are important. As a result, the set of ideal conditions, $\B\cdot \E=0$ and $B\geq E$, is violated. Since the approach we took - expansion in small velocity - involves step-by-step approximation, it is feasible that higher order effects will smooth-out the divergencies. We think this is unlikely: divergent first-order \Ef\ is not compensated by the second order \Bf, Eq. (\ref{E1overB2}). Instead, the second order \Bf\ is divergent on its own. Divergent electric currents,  Eq. (\ref{JEM}) will lead to resistive dissipation.

\begin{figure}  [h!]
    \includegraphics[width=.99\linewidth]{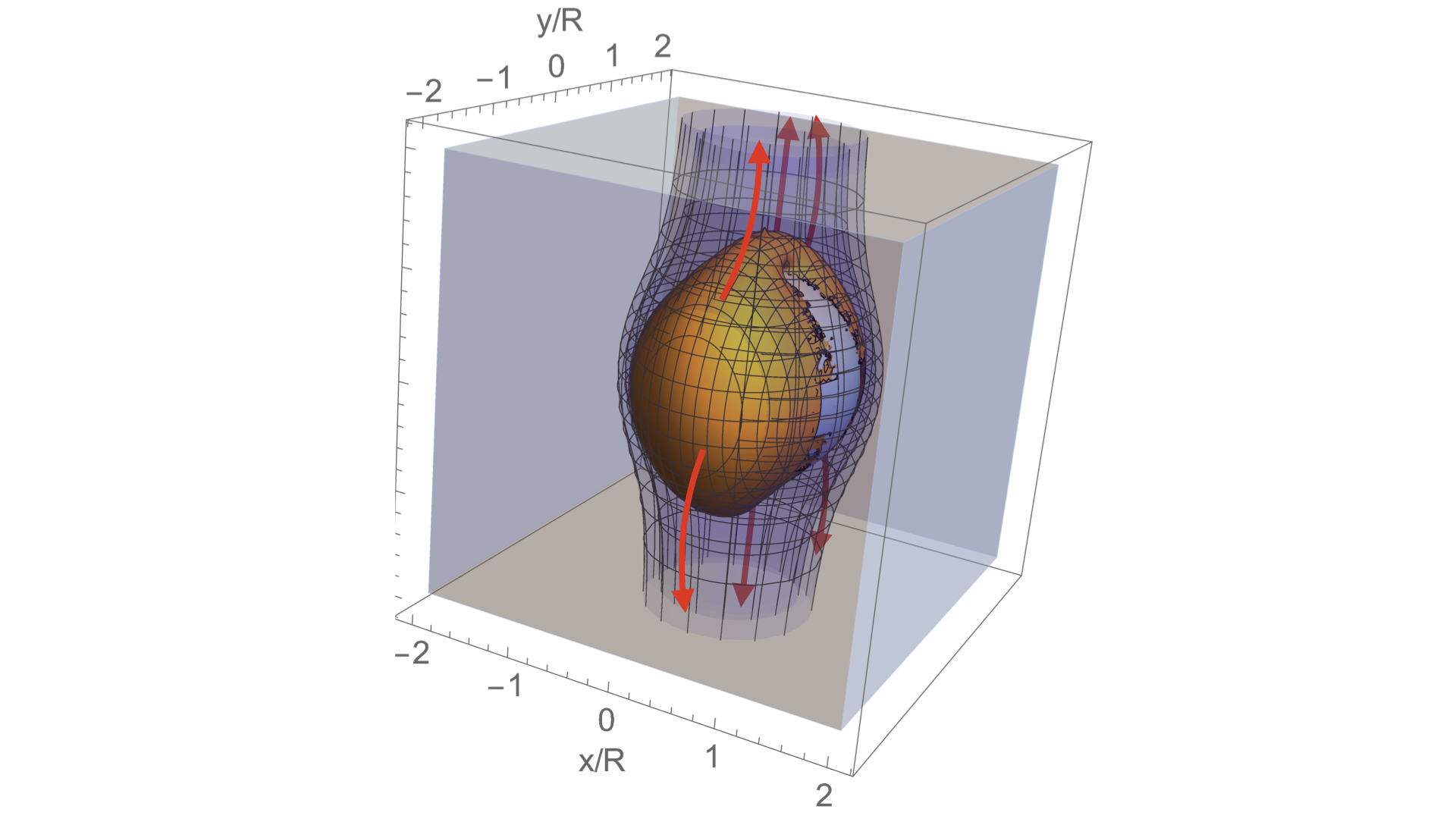}
\caption{Expected jets from a \NS\ moving through force-free \Bf. Yellow regions are dissipative regions, $E\geq B$. Quasi-cylindrical surfaces are magnetic flux surfaces. Dissipation within the $E\geq B$  regions would produce   double-tongue-like jet  structures. } 
\label{outline}
\end {figure}

Thus, we expect \EM\ dissipation near the \NS\ (or magnetospheric boundary).  Particle will be accelerated and eventually collimated to move  that  particle   along \Bf\ lines, Fig. \ref{outline}.  

The effect of collimation may be important for the detection of precursors, since the expected powers are not very high.
The expected powers  in the 1M-DNS and 2M-DNS scenarios were discussed by  \cite{2019MNRAS.483.2766L}.
If a \NS\ is moving in the field of a primaries' dipolar \Bf\ at orbital separation  $r$, 
the expected  powers is \citep{2001MNRAS.322..695H,2011PhRvD..83l4035L}
\be
L_1 \sim  \frac{G B_{{NS}}^2 M_{{NS}} R_{{NS}}^8}{c r^7}=3 \times { 10^{41} }{(-t)^{-7/4}}\, {\rm \, erg \, s^{-1}}
\label{L1}
\ee
where in the last relations the time to merger $t$ is measured in seconds.
(Index $1$ indicates here that the interaction is between single magnetized \NS\ and unmagnetized one.) 
 Magnetospheric interaction of two magnetized \NSs\ can generate larger luminosity that the case of  one star moving in the field of the companion  \citep{2019MNRAS.483.2766L}. In this case 
 \be L_2 \sim \frac{ B_{{NS}}^2 G  M_{{NS}} R_{{NS}}^6}{c  r^5}    =
\frac{c^{21/4} B_{{NS}}^2 R_{{NS}}^6}{
   (-t)^{5/4} ( G M_{NS})^{11/4}}=
 6 \times { 10^{42} }{(-t)^{-5/4}}\, {\rm \, erg \, s^{-1}}
    \label{L2}
\ee
(Index $2$ indicates here that the interaction is between two magnetized \NS.)
The ratio of luminosities of the models  1M-DNS  and 2M-DNS is 
\be
\frac{L_2}{L_1}  = \left( \frac{G M}{c^2 R_{NS}}\right)^{3/2} \sqrt{\frac{(- t) c}{R_{NS}}} \approx 16 \sqrt{-t}
\ee
Thus $L_2$ dominates $L_1$ prior to merger. This is due to larger interaction region, of the order of teh magnetospheric radius, instead of the radius of a \NS.

Qualitatively, for the non-magnetar \Bf\  the power (\ref{L2}) is fairly small.  Even at the time of a merger, with $t\sim 10^{-2}$   seconds the corresponding power is only
  $L \sim  10^{45}{\rm \, erg \, s^{-1}}$ - hardly observable  from cosmological distances by all-sky monitors.  The best case is if a fraction of the power (\ref{L2}) is put into radio.  If a fraction of $\eta_R$ of the power is put into radio, the expected signal then is 
\be
F_R \sim \eta_R  \frac{L_2}{4\pi d^2  \nu}\approx 0.1 {\rm \, Jy}\,  \eta_{R,-5} (-t)^{-5/4}
\label{FR}
\ee
This is a fairly strong signal that could be detected by  modern radio telescopes. 

Our results indicate that even in the single magnetized case we expect relativistic jets  (or ``tongues'') produced due to \EM\ interaction of merging \NSs, with correspondingly beamed emission pattern. Another way to produce higher luminosity is at the moments of topological spin-orbital resonances \citep{2021ApJ...923...13C}.

\section*{Acknowledgments}
   This work had been supported by  NSF grants 1903332 and  1908590.


\section{Data availability}
The data underlying this article will be shared on reasonable request to the corresponding author.

\bibliographystyle{apj}

 \bibliography{/Users/maxim/Dropbox/Research/BibTex}

\appendix

\section{1st order variation of $\alpha-\beta-\Phi$ are vanishing}
\label{vanishing}
Here we demonstrate that in the first order of $v_0$ the   variation of $\alpha-\beta-\Phi$ are vanishing.

Let us expand
\ba &&
 \alpha= \alpha_0+ \epsilon  \alpha_1
 \nn && 
\beta=\beta_0 +  \epsilon  \beta_1
\nn &&
\Phi = \Phi_0 +   \epsilon  \Phi_1
\ea

The orthogonality constraint 
\be
(\nabla \alpha ) \cdot (\nabla \beta )=0 \to (\nabla \alpha_0 ) \cdot (\nabla \beta_1) + (\nabla \alpha_1 ) \cdot (\nabla \beta_0) =0
\ee
implies
\ba &&
\nabla \alpha _1= a_1 \nabla \alpha_0 + a_2 \nabla \Phi _0
\nn && 
\nabla \beta_1 =  b_1 \nabla \beta_0 + b_2 \nabla \Phi_0
\ea

On the other hand, 
\be
(\nabla \alpha ) \cdot (\nabla \Phi )=0 \to (\nabla \alpha_0 ) \cdot (\nabla \Phi_1) + (\nabla \alpha_1 ) \cdot (\nabla \Phi_0) =0
\ee
hence
\ba &&
\nabla \alpha _1= a_1 \nabla \alpha_0 + d_2 \nabla \beta_0
\nn && 
\nabla \Phi_1 =  c_1 \nabla \Phi_0 + c_2 \nabla \beta_0
\ea

Thus, to keep all surfaces orthogonal we need
\ba &&
\nabla \alpha _1= a_1 \nabla \alpha_0
\nn &&
\nabla \beta_1 =  b_1 \nabla \beta_0 
\nn &&
\nabla \Phi_1 =  c_1 \nabla \Phi_0
\ea

Thus, first order perturbations are ``locked in''.

\section{Comparing \Ef\ (\ref{EE1})  with other cases}
\label{comp}

The \Ef\ (\ref{EE1}) is not too different from the vacuum case, where for \Ef\ along $y$ direction at infinity
    \ba && 
     \Phi^{(vac)}= E_0 (1-(R/r)^3) r \sin \theta \sin \Phi
     \nn &&
     E_r ^{(vac)}= -(1+ 2 R^3/r^3) \sin \theta \sin \Phi E_0
     \nn &&
     E_\theta^{(vac)}= (1- (R/r)^3) r \cos \theta \sin \Phi E_0
     \nn && 
     E_\phi^{(vac)} = (1-(R/r)^3) \cos \Phi
     \label{EE1vac}
     \ea 
with surface charge density     
     \be
     \sigma^{(vac)} = \frac{3}{2\pi} \sin  \Phi \sin \theta E_0
     \ee
     Fields   (\ref{EE1})  and   (\ref{EE1vac})  have the same angular dependence, but different radial dependence.
The \Ef\  (\ref{EE1vac}) has a non-zero component along $B_0$. 
\be
  \E ^{(vac)} \cdot \B_0 =  \frac{3 R^3   \left(r^3-R^3\right)\sin (2 \theta ) \sin \phi}{4 r^6} E_0 B_0
\ee

Another  possible approximation, that of an  incompressible flow around a sphere  with kinematically added \Bf, with velocity
\be
{\bf v}^{(inc)} = \left\{ -(1-R^3/r^3)  \sin \theta \cos \phi  , - \left(1+ R^3/(2 r^3) \right)  \cos \theta \cos \phi , \left(1+ R^3/ r^3 \right)  \sin\phi  \right\} v_0
  \ee
  would produce \Ef\ with similar angular dependence,
  \ba &&
  \E ^{(inc)} = - {\bf v}^{(inc)}  \times \B_0= 
  \nn &&
  E_r = \left( 1+ R^3/(2 r^3) \right)  ^2 \sin \theta \sin \phi  \, v_0  B_0
  \nn &&
  E_\theta =  \left( 1- R^3/(2 r^3) -   R^6/(2 r^6)\right)  \cos \theta \sin \phi  \,  v_0  B_0
  \nn &&
  E_\phi =  \left( 1- R^3/(2 r^3) -   R^6/(2 r^6)\right) \cos \phi \, 
v_0 B_0
   \ea
A drawback of this approach is that the \Ef\  has  non-zero curl
   \be
   \nabla \times  \E ^{(inc)} =\left\{0,-\frac{9 R^6 \cos \phi}{4 r^7},\frac{9 R^6 \cos  \theta  \sin \phi}{4 r^7}\right\}
    v_0 \B_0
    \ee
    and hence cannot be stationary.

\end{document}